\documentclass[journal, onecolumn]{IEEEtran}

\usepackage{cite}

\ifCLASSINFOpdf
   \usepackage[pdftex]{graphicx}
\else
\fi
\usepackage{amsmath}
\usepackage{algorithmic}

\usepackage{array}

\usepackage{algorithm}
\begin{document}
%
\title{BGD-based Adam algorithm for time-domain equalizer in PAM-based optical interconnects}
%
%
%

\author{Haide Wang$^1$,
        Ji Zhou$^{1,*}$,
        Weiping Liu$^1$,
        Jianping Li$^2$,
        Xincheng Huang$^1$,
        Long Liu$^1$,
        Weixian Liang$^1$,
        Changyuan Yu$^3$,
        Fan Li$^4$,
        and~Zhaohui Li$^{4,5}$
\thanks{$^1$Department of Electronic Engineering, College of Information Science and Technology, Jinan University, Guangzhou 510632, China.}
\thanks{$^2$Provincial Key Laboratory of Optical Fiber Sensing and Communications, Institute of Photonics Technology, Jinan University, Guangzhou 510632, China.}
\thanks{$^3$Department of Electronic and Information Engineering, Hong Kong Polytechnic University, Hong Kong, China.}
\thanks{$^4$State Key Laboratory of Optoelectronic Materials and Technologies, School of Electronics and Information Technology, Sun Yat-sen University, Guangzhou, 510275, China.}
\thanks{$^5$e-mail:lzhh88@mail.sysu.edu.cn.}
\thanks{$^*$Corresponding author: zhouji@jnu.edu.cn.}}

\maketitle

\begin{abstract}
To the best of our knowledge, for the first time, we propose adaptive moment estimation (Adam) algorithm based on batch gradient descent (BGD) to design a time-domain equalizer (TDE) for PAM-based optical interconnects. Adam algorithm has been widely applied in the fields of artificial intelligence. For TDE, BGD-based Adam algorithm can obtain globally optimal tap coefficients without being trapped in locally optimal tap coefficients. Therefore, fast and stable convergence can be achieved by BGD-based Adam algorithm with low mean square error. Meanwhile, BGD-based Adam algorithm is implemented by parallel processing, which is more efficient than conventional serial algorithms, such as least mean square and recursive least square algorithms. The experimental results demonstrate that BGD-based Adam feed-forward equalizer works well in 120-Gbit/s PAM8 optical interconnects. In conclusion, BGD-based Adam algorithm shows great potential for converging the tap coefficients of TDE in future optical interconnects.
\end{abstract}

%
\IEEEpeerreviewmaketitle

\section{Introduction}
%
%
%
%
\IEEEPARstart{O}{wing} to the emergence of cloud computing and a variety of web applications, large-scale data centers nowadays are resorting to optical interconnects to meet the explosive increase of network traffic \cite{cheng2018recent}. To achieve higher capacity, eight-level pulse-amplitude modulation (PAM8) is a potential for future optical interconnects although it is sensitive to inter-symbol interference (ISI) and noise \cite{chen2018nonlinear, diamantopoulos2019amplifierless}. In general, time-domain equalizer (TDE) can be employed to compensate ISI for PAM8 system. Recursive least squares (RLS) and least mean squares (LMS) algorithms are two common adaptive algorithms to converge the tap coefficients for TDE. However, with the increase of data rate and modulation level, TDE using RLS algorithm has high computational complexity and TDE using LMS algorithm requires large amount of training samples, which may be not well-suited for future optical interconnects \cite{zhang2016single}.

In recent years, machine learning (ML) algorithms have been widely applied in the fields of artificial intelligence (AI) \cite{jordan2015machine}. Since ML algorithms require large scale training data sets, many efficient adaptive algorithms have been proposed for fast and stable convergence to minimize the error function of ML algorithms \cite{duchi2011adaptive, zeiler2012adadelta, kingma2014adam}. Since it¡¯s believed that more data beats better algorithm in AI fields \cite{domingos2012few}, AI scientists often use the distinguished adaptive moment estimation (Adam) algorithm for stochastic optimization without reducing the scale of training set. Generally, there is a trade-off between the accuracy and the number of training samples in the training process \cite{bottou2008tradeoffs}. It¡¯s very possible to optimize the TDE with traditional structure by using these adaptive algorithms. However, different from ML, in real communication systems, we are supposed to use as few training samples and low computational complexity as possible to get the optimal tap coefficients.

In this paper, inspired by advances of AI, we first propose batch gradient descent (BGD)-based Adam algorithm to achieve fast and stable convergence of tap coefficients in feed-forward equalizer (FFE), a major type of TDE. As we known, the conventional RLS and LMS algorithms employ the serial processing to converge the tap coefficients. However, BGD-based Adam algorithm for FFE can be implemented by an effective parallel processing. Meanwhile, BGD-based Adam algorithm has adaptive step size for realizing precise convergence with low mean square error (MSE). The experimental results of 120-Gbit/s PAM8 optical interconnects demonstrate that BGD-based Adam FFE can effectively get the optimal tap coefficients using less training samples and iterations, which shows great potential for future optical interconnects.

\section{Princeple of BGD-based Adam algorithm for FFE}
BGD-based Adam algorithm for FFE requires $M$ training samples to perform the update of tap coefficients in the training process. At the receiver, the training samples are received and stored to construct a training matrix $\emph{\textbf R}$ for parallel processing. During the procedure of training, all the $M$ samples are required to be stored. The structure of the training matrix $\emph{\textbf R}$ can be expressed as
\begin{equation}
\begin{aligned}
\emph{\textbf R}&=\left[ \begin{array}{cccc}
{x_{N}} & {x_{N-1}}& {\dots} & {x_{1}} \\ {x_{N+1}} & {x_{N}}& {\dots} & {x_{2}} \\ {\vdots} & {\vdots}& {\ddots} & {\vdots} \\ {x_{M}} & {x_{M-1}}& {\dots} & {x_{M-N+1}}\end{array}\right] = \left[ \begin{array}{c} \emph{\textbf r}_N \\ \emph{\textbf r}_{N+1} \\ {\vdots} \\ \emph{\textbf r}_{M}\end{array}\right]
\end{aligned}
\label{eq:refname1}
\end{equation}
where $x$ are the received training samples. Obviously, the dimension of $\emph{\textbf{R}}$ is $(M - N + 1)$-by-$N$ where $M$ is the number of training samples and $N$ is the number of taps in FFE. The transmitted training vector is
\begin{equation}
\begin{aligned}
\emph{\textbf Y}
&=[ \begin{array}{cccc}{y_{N}} &{y_{N+1}} &{\dots}  &{y_{M}}\end{array}]^{T}
\end{aligned}
\label{eq:refname2}
\end{equation}
where $(\cdot)^{T}$ denotes matrix transpose. The error function used in BGD-based Adam algorithm for FFE is MSE, which can be expressed as
\begin{equation}
\begin{aligned}
J(\boldsymbol{\omega})&=\frac{1}{M-N+1} \sum_{i=N}^{M}\left(\emph{\textbf r}_{i}  \boldsymbol{\omega}-y_{i}\right)^{2}\\
&=\frac{1}{M-N+1}(\emph{\textbf R}  \boldsymbol{\omega}-\emph{\textbf Y})^{T}(\emph{\textbf R}  \boldsymbol{\omega}-\emph{\textbf Y})
\end{aligned}
\label{eq:refname3}
\end{equation}
where $\boldsymbol{\omega}=[\begin{array}{llll}{\omega_{1}} & {\omega_{2}} & {\dots} & {\omega_{N}}\end{array}]^{T}$ is the tap coefficient vector of the FFE. Gradient $\emph{\textbf G}$ is the partial derivative of $J(\boldsymbol{\omega})$ with respect to $\boldsymbol{\omega}$, which can be calculated as
\begin{equation}
\begin{aligned}
\emph{\textbf G} &=[ \begin{array}{cccc}{\frac{\partial J(\boldsymbol{\omega})}{\partial \omega_{1}}} &{\frac{\partial J(\boldsymbol{\omega})}{\partial \omega_{2}}}  &{\dots} &{\frac{\partial J(\boldsymbol{\omega})}{\partial \omega_{N}}}\end{array}]^{T}\\
&= \frac{2}{M-N+1} \times \left[ \begin{array}{c}\sum_{i=N}^{M}\left(\emph{\textbf r}_{i}  \boldsymbol{\omega}-y_{i}\right) * x_{i} \\\sum_{i=N}^{M}\left(\emph{\textbf r}_{i}  \boldsymbol{\omega}-y_{i}\right) * x_{i-1} \\{\vdots} \\\sum_{i=N}^{M}\left(\emph{\textbf r}_{i}  \boldsymbol{\omega}-y_{i}\right) * x_{i-N+1}\end{array}\right] \\
&= \frac{2}{M-N+1} \emph{\textbf R}^{T} (\emph{\textbf R}  \boldsymbol{\omega}-\emph{\textbf Y}).
\end{aligned}
\label{eq:refname4}
\end{equation}

\begin{algorithm}[!t]
\caption{BGD-based Adam algorithm for FFE}\label{alg:Adam}
\begin{algorithmic}[1]
\REQUIRE~~ $\emph{\textbf Y}$ \COMMENT{Training vector}
\REQUIRE~~ $\emph{\textbf R}$ \COMMENT{Received training matrix}
\REQUIRE~~ $I$ \COMMENT{Total iteration number}
\REQUIRE~~ $\theta$ \COMMENT{Step size}
\REQUIRE~~ $\boldsymbol{\omega} = zeros(N,1)$ \COMMENT{Initialize tap coefficients}
\REQUIRE~~ $\textbf{m}_0 = zeros(N,1)$ \COMMENT{Initialize first moment vector}
\REQUIRE~~ $\textbf{v}_0 = zeros(N,1)$ \COMMENT{Initialize second moment vector}
\REQUIRE~~ $t = 0$ \COMMENT{Iteration initialization}
\STATE $t = t+1$
\STATE $\emph{\textbf G}_{t} =  \frac{2}{M-N+1} \times\emph{\textbf R}^T(\emph{\textbf R} \boldsymbol{\omega}_{t-1} - \emph{\textbf Y})$ \COMMENT{Get gradients}
\STATE $\textbf{m}_t = \beta_1\times \textbf{m}_{t-1}+(1-\beta_1)\times \emph{\textbf G}_{t}$
\STATE $\textbf{v}_t = \beta_2\times \textbf{v}_{t-1}+(1-\beta_2)\times \emph{\textbf G}^2_t$
\STATE $\hat{\textbf{m}}_t = \textbf{m}_{t}/(1-\beta^t_1)$ \COMMENT{Bias-corrected operation}
\STATE $\hat{\textbf{v}}_t = \textbf{v}_{t}/(1-\beta^t_2)$ \COMMENT{Bias-corrected operation}
\STATE $\boldsymbol{\omega_{t}} = \boldsymbol{\omega}_{t-1}-\theta\times \hat{\textbf{m}}_t/ (\sqrt{\hat{\textbf{v}}_t} + \epsilon)$ \COMMENT{Update tap coefficients}
\STATE \textbf{end while}
\STATE \textbf{return} $\boldsymbol{\omega}_{I}$ \COMMENT{Return tap coefficients}
\end{algorithmic}
\end{algorithm}

Conventional BGD method updates $\boldsymbol{\omega}$ in the opposite direction of the gradient $\emph{\textbf G}(\boldsymbol{\omega})$, which can be expressed as
\begin{equation}
\boldsymbol{\omega}_{t} =\boldsymbol{\omega}_{t-1}-\theta \times \emph{\textbf G}_{t-1}
\label{eq:refname5}
\end{equation}
where $\theta$ is a fixed step size ranging from $0$ to $1$ and subscript $t$ denotes $t$-th iteration. Generally speaking, when the step size is too large, it may fail to converge, or even diverge; but it needs a great number of iterations when the step size is too small \cite{goodfellow2016deep}. However, BGD-based Adam algorithm is much less sensitive to the step size $\theta$ compared with conventional BGD method for the reason that it computes adaptive step sizes from estimates of biased first and second moments of gradients. BGD-based Adam algorithm for FFE in the training process is illustrated in Algorithm \ref{alg:Adam}. The biased first and second moment estimates  $\mathbf{m}_{t}$ and $\mathbf{v}_{t}$ of $\emph{\textbf G}_t$ are initialized as zeros vector, which can be expressed as
\begin{equation}
\mathbf{m}_{t}=\beta_{1} \times \mathbf{m}_{t-1}+\left(1-\beta_{1}\right) \times \emph{\textbf G}_{t},
\label{eq:refname6}
\end{equation}
\begin{equation}
 \mathbf{v}_{t}=\beta_{2} \times \mathbf{v}_{t-1}+\left(1-\beta_{2}\right) \times \emph{\textbf G}_{t}^{2}
\label{eq:refname7}
\end{equation}
where $\beta_{1}$ and $\beta_{2}$ are set to 0.9 and 0.999, respectively. The bias-corrected operations keep the biased first and second moment estimates from moving towards zeros at the beginning of iterations, which can be expressed as
\begin{equation}
\hat{\mathbf{m}}_{t} =\mathbf{m}_{t} /\left(1-\beta_{1}^{t}\right),
\label{eq:refname8}
\end{equation}
\begin{equation}
\hat{\mathbf{v}}_{t} =\mathbf{v}_{t} /\left(1-\beta_{2}^{t}\right).
\label{eq:refname9}
\end{equation}
A relative small value $\epsilon$ is used to prevent zero-division error and the tap coefficients are updated as
\begin{equation}
\boldsymbol{\omega_{t}} = \boldsymbol{\omega}_{t-1}-\theta\times \hat{\textbf{m}}_t/ (\sqrt{\hat{\textbf{v}}_t} + \epsilon).
\label{eq:refname10}
\end{equation}

BGD-based Adam algorithm calculates the error function after scanning all training samples and then updates parameters. It¡¯s acknowledged that BGD-based Adam, which requires all training samples every iteration, is guaranteed to converge to globally optimal solution for convex error function, such as MSE function. However, other gradient descent methods which do not use all training samples every iteration, are more likely to be trapped in the locally optimal coefficients and frequent updating may result in drastic fluctuation of the error function \cite{ruder2016overview}. Moreover, BGD-based Adam algorithm updates the parameters also less frequently than LMS and RLS algorithms, which uses only one training sample every iteration and run in serial. However, it should be noted that BGD-based method needs an extra memory to store all training samples in the training process. It's worth noting that a significant feature of BGD-based Adam algorithm is that the basic operations are based on matrices and vectors. It means that it runs much faster in parallel in some computing environments, such as MATLAB, Numerical Python in AI field, field programmable gata array in industrial field and so on \cite{van2011numpy, dou200564}.

Further, after converging to the globally optimal tap coefficients by BGD-based Adam algorithm, the extra memory aren't needed and the system serially equalizes the received signals. After equalization, a simple post filter is used to suppress the amplified high-frequency noise. The output of the post filter can be express as
\begin{equation}
z_{k}=s_{k}+\alpha \times s_{k-1}
\label{eq:refname11}
\end{equation}
where $s_{k}$ is the output of the FFE, and $\alpha$ is the tap coefficient of post filter. Further, the post filter unavoidably introduces a known ISI, but it can be eliminated by maximum likelihood sequence detection (MLSD) algorithm \cite{zhong2018digital}.

\section{Experimental setups}
\begin{figure}[!t]
\centering
\includegraphics[width=4in]{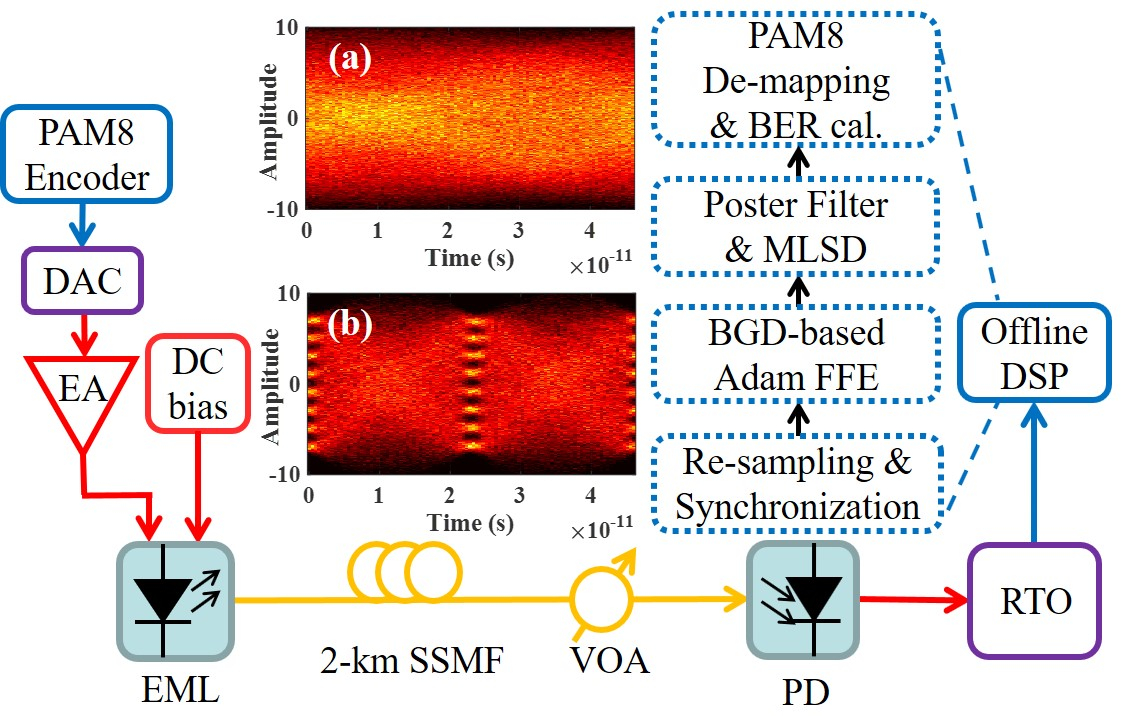}
\caption{Experiment setups. EML, electro-absorption modulator integrated laser; DAC, digital-to-analog converter; EA, electric amplifier; DC bias, direct current bias; SSMF, standard single-mode fiber; VOA, variable optical attenuator; PD, photodiode; RTO, real-time oscilloscope. Inset (a) is eye diagram of received PAM8 signals. Inset (b) is equalized eye diagram of PAM8 signals.}
\label{fig:exsetup}
\end{figure}

\begin{figure*}[!b]
\centering
\includegraphics[width=6.2IN]{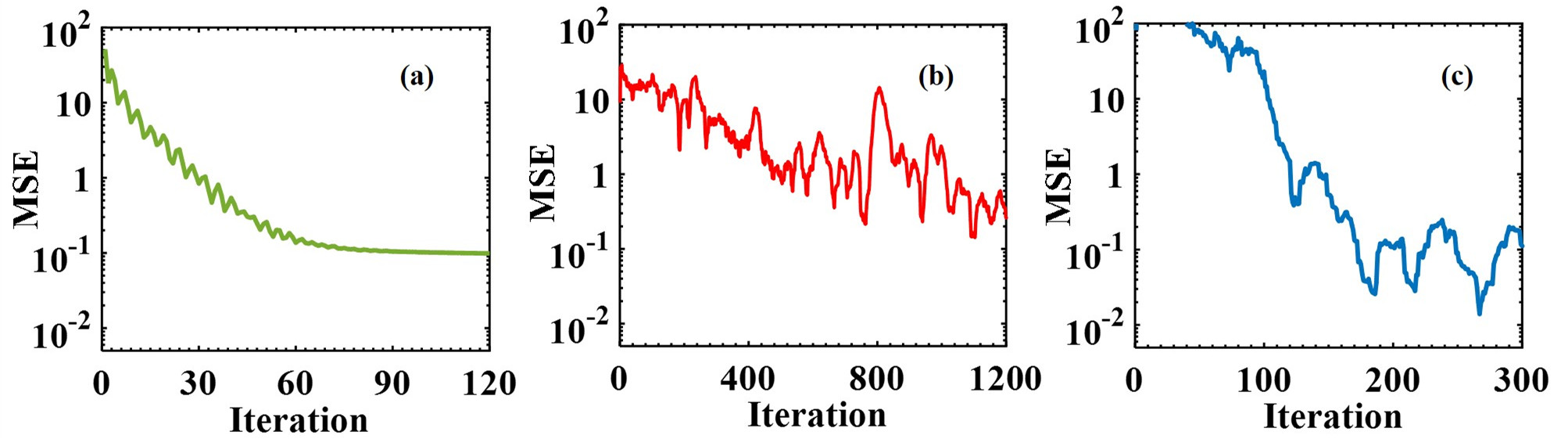}
\caption{MSE curves of FFE applied for 120-Gbit/s PAM8 optical interconnects. (a) BGD-based Adam, (b) LMS, (c) RLS algorithm are employed, respectively.}
\label{fig:MSE}
\end{figure*}

Fig. \ref{fig:exsetup} shows the experimental setups of 120-Gbit/s PAM8 system using BGD-based Adam FFE. At the transmitter, the digital frames of PAM8 signals are uploaded into a digital-to-analog converter (DAC) with 86-GSa/s sampling rate and 16-GHz 3-dB bandwidth to generate electrical PAM8 frames. The symbol rate of electrical PAM8 frames is set to 43 GBaud. After being amplified by an electrical amplifier (EA), the amplified electrical PAM8 frames are modulated by a 40-Gbit/s electro-absorption integrated laser modulator (EML), to which an appropriate direct current (DC) bias is applied, to generate the optical PAM8 frames. Subsequently, the generated optical PAM8 signals are launched into 2-km standard single mode fiber (SSMF). At the receiver, a variable optical attenuator (VOA) is employed to adjust the received optical power (ROP) of signals. Then a photodiode (PD) converts the received optical signals into electrical signals. The electrical signals are converted into digital signals by a real-time oscilloscope (RTO) with sampling rate of 80 GSa/s and 3-dB bandwidth of 36 GHz. Finally, off-line processing is implemented to deal with the digital signals, including re-sampling, synchronization, BGD-based Adam FFE, post filter, MLSD, PAM8 de-mapping and bit error rate (BER) calculation. When the lengths of training samples and total samples in one frame is set to 300 and 164480, respectively, the net rate of electrical PAM8 frames is approximately 120 Gbit/s ($3\times 43\times 164180/164480/(1+7\%) \approx 120$ Gbit/s). The eye diagrams of received PAM8 signals and equalized PAM8 signals are shown as Inset (a) and (b), respectively. Apparently, the serious ISI is effectively compensated after equalization.

\section{Results and discussion}
The curves of MSE of FFE with BGD-based Adam, LMS and RLS algorithm are shown in Fig. \ref{fig:MSE}. As shown in Fig. \ref{fig:MSE} (a), on the one hand, after 100 iterations, MSE of BGD-based Adam algorithm is around $0.1$ and obviously it has converged. Although computational complexity of each iteration of BGD-based Adam algorithm is higher than that of the other two algorithms, it updates the coefficients of tap $\boldsymbol{\omega}$ with all training samples every iteration less frequently and steadily. As shown in Fig. \ref{fig:MSE} (b) and Fig. \ref{fig:MSE} (c), it's clear that LMS or RLS algorithm cannot converge even after 200 iterations and the MSE curves are much more fluctuated than that of BGD-based Adam algorithm for the reason that LMS or RLS algorithm frequently updates the tap coefficients giving rise to MSE with a high variance. BER performances of the above algorithms undoubtedly become better with the increase of iterations.

\begin{figure}[t]
\centering
\includegraphics[width=6.5 IN]{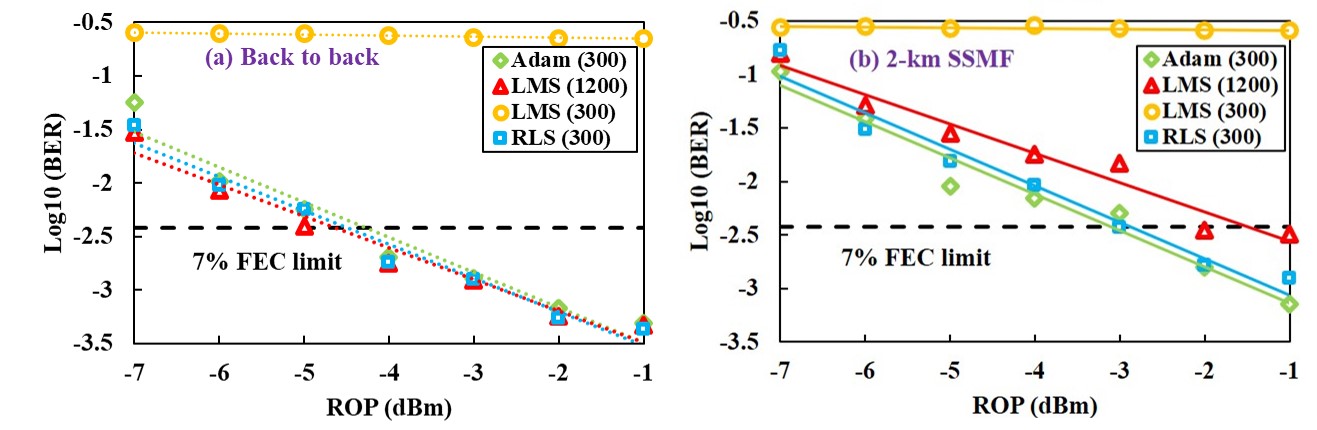}
\caption{BER performances of 120-Gbit/s PAM8 optical interconnects versus ROPs at BTB (a) and 2-km transmission (b) with FFE, post filter and MLSD. FFE employs BGD-based Adam with 300 training samples (green rhombus), LMS with 1200 (red triangle) or 300 (yellow circle) training samples, and RLS with 300 training samples (blue square), respectively.}
\label{fig:BER}
\end{figure}

As shown in Fig. \ref{fig:BER} (a), for 120-Gbit/s PAM8 system after back-to-back (BTB) transmission, BGD-based Adam algorithm using 300 training samples has almost the same performance as those of RLS algorithm with 300 training samples and LMS algorithm with 1200 training samples. However, LMS algorithm doesn't work well for converging the tap coefficients when the training samples are set to 300. Therefore, BGD-based Adam algorithm is more efficient than LMS algorithm. As shown in Fig. \ref{fig:BER} (b), BGD-based Adam algorithm using 300 training samples can achieve good and stable performances for 120-Gbit/s PAM8 system after 2-km SSMF transmission, which is comparable to RLS algorithm using 300  training samples but better than LMS algorithm using 1200 training samples. The ROP of 120-Gbit/s PAM8 system with BGD-based Adam algorithm is approximately 1-dB lower than that with LMS algorithm at the 7\% FEC limit. Therefore, BGD-based Adam algorithm is more robust for resisting the limited bandwidth and noise than LMS algorithm even using less training samples. The tap numbers of FFE using the above algorithms are set to 181.

The computational complexity of the training processes can be calculated as \cite{zhou2019joint}
\begin{equation}
\mathrm{C}_{\text {BGD-Adam}} = N\times[2(M-N+1)+9]\times I,
\label{eq:refname12}
\end{equation}
\begin{equation}
\begin{array}{c}\mathrm{C}_{\text {LMS}} = (2N+1)\times M,~\text{and}
\end{array}
\label{eq:refname13}
\end{equation}
\begin{equation}
\begin{array}{c}\mathrm{C}_{\text {RLS}} = (3N^{2}+5 N +2)\times M
\end{array}
\label{eq:refname14}
\end{equation}
where $M$ is the length of training sequences, $N$ is the number of taps and $I$ is the iteration number of BGD-based Adam algorithm. $I$ and $N$ are respectively set to 120 and 181. The computational complexity of BGD-based Adam algorithm is a concave quadratic function of the tap number $N$, while the computational complexity of LMS algorithm is proportional to $N$ and  the computational complexity of RLS algorithm is proportional to $N^{2}$.

Table \ref{tab:1} shows the comparisons of training samples, computational complexity and run mode of BGD-based Adam, LMS and RLS algorithms in the training process. Owing to the precise convergence of batch training, the training samples of BGD-based Adam algorithm is the same as RLS algorithm and 75\% less than that of LMS algorithm. In general, the computational complexities of the mentioned adaptive algorithms in the training process are shown in ascending order as follows: $\mathrm{C}_{\text{LMS}}<\mathrm{C}_{\text{BGD-Adam}}<\mathrm{C}_{\text {RLS}}$. The computational complexity of BGD-based Adam algorithm is higher than that of LMS, but less than that of RLS algorithm. However, thanks to the adaptive step size, the tap coefficients converge rapidly and steadily by using BGD-based Adam algorithm. Therefore, the iteration number of BGD-based Adam algorithm is usually smaller than that of LMS and RLS algorithm. Furthermore, parallel computation techniques greatly speed up BGD-based Adam algorithm.

\begin{table}[!t]
\centering
\caption{\bf Comparisons of number of training samples, computational complexity and run mode of BGD-based Adam, LMS and RLS algorithms in the training process.}

\begin{tabular}{c|ccc}
\hline
&$M$ &$\mathrm{C}_\text{Training}$ &Run mode\\
\hline
BGD-based Adam &300 &$\sim 10^{6}$ &Parallel\\
\hline
LMS &1200 &$\sim 10^{5}$ &Serial\\
\hline
RLS &300 &$\sim 10^{7}$ &Serial\\
\hline
\end{tabular}
\label{tab:1}
\end{table}

\section{Conclusion}
In conclusion, for the first time, we propose BGD-based Adam TDE for PAM-based optical interconnects. The experimental results of 120-Gbit/s PAM8 optical interconnects over 2-km transmission demonstrate that BGD-based Adam TDE can effectively and efficiently get the optimal tap coefficients using less training samples and iterations to approach good performance. BGD-based Adam algorithm achieves a better performance than LMS algorithm. Furthermore, the computational complexity of BGD-based Adam algorithm is lower than RLS algorithm and it can be accelerated owing to the matrix operations in parallel. BGD-based Adam algorithm is robust and suitable for resisting the limited bandwidth and serious noise, showing great potential for future optical interconnects.


%

\section*{Funding.} The Science and Technology Planning Project of Guangdong Province (2017B010123005, 2018B010114002). Local Innovation and Research Teams Project of Guangdong Pearl River Talents Program (2017BT01X121); National Science Foundation of China (NSFC) (61525502); The Fundamental Research Funds for the Central Universities (21619309).

\ifCLASSOPTIONcaptionsoff
  \newpage
\fi



%

\end{document}